\documentclass[final,5p,times,twocolumn]{elsarticle}

\usepackage{amssymb}

\usepackage{lineno}
\usepackage{enumitem} 
\usepackage{url}
\usepackage{graphicx}
\usepackage{subfig}
\usepackage{amsmath}
\usepackage{booktabs}
\usepackage[usenames,dvipsnames]{color}

\journal{Nuclear Instruments and Methods A}

\begin{document}

\begin{frontmatter}



\title{Evaluation of the absolute single-photon detection efficiency of HRPPD}


\author[BNL]{Yifan Jin}
\author[BNL]{Alexander Kiselev}
\author[BNL]{Sean Stoll}

\address[BNL]{Physics Department, Brookhaven National Laboratory, Upton, NY 11733, USA.}

\begin{abstract}
    \noindent Pixelated High Rate Picosecond Photon Detectors (HRPPDs) by Incom Inc. are promising photosensors for use in Ring Imaging CHerenkov (RICH) detectors, where a high gain, sub-mm position resolution and sub-100ps timing resolution are required in a single photon mode. Quantum Efficiency (QE) has been measured for the first batch of EIC HRPPDs both by Incom and EIC research groups at Jefferson Lab and Brookhaven Lab, with peak values at $\sim$365~nm typically exceeding 30\%. In this study, we present a first direct measurement of Photon Detection Efficiency being equal to (17.1 $\pm$ 0.1 [stat] $\pm$ 0.3 [sys])\% at 398.6~nm, for a pixel near the center of HRPPD, in a photoelectron pulse counting mode using a picosecond diode laser. HRPPD QE at the same spot and at the same wavelength 
    was evaluated to be (24.4 $\pm$ 0.1 [stat] $\pm$ 0.3 [sys])\%, leading to a Collection Efficiency estimate of (70.3 $\pm$ 1.6)\%, which is consistent with the expectations.  
\end{abstract}


\end{frontmatter}



\section{Introduction}
Pixelated High Rate Picosecond Photon Detectors (HRPPDs), a new type of microchannel plate photomultiplier tube, thanks to their high gain, excellent timing resolution, and sub-mm position resolution, serve as promising candidates for use in Ring Imaging CHerenkov (RICH) detectors at the future EIC~\cite{EIC} and upgrades for LHCb~\cite{CERN-LHCC-2021-012} and Belle II~\cite{belleII}. 

In contrast to Large Area Picosecond Photo-Detectors~\cite{LAPPD1, LAPPD2, LAPPD3, LAPPD4, LAPPD5, LAPPD6}, which either employ stripline anodes for direct charge readout through hermetic feedthroughs (Gen-I), or employ a uniform ceramic anode plate with a resistive layer capacitively coupled with an external readout board (Gen-II), HRPPDs utilize pixellated anodes with direct charge collection, similar to Planacon MCP-PMT~\cite{Burle2004, Burle2005}.

Recently, Quantum Efficiency (QE) of the first batch of EIC HRPPDs produced by Incom Inc.~\cite{Incom} has been measured both by Incom and EIC research groups at Jefferson Lab and Brookhaven Lab, with peak values at $\sim$365~nm typically exceeding 30\%~\cite{Lyashenko}. In addition to QE, the photon detection efficiency (PDE) provides a more direct and practical measure of the device's feasibility for use in RICH detectors.
In this paper, we present the results of the first measurement of the absolute photon detection efficiency of HRPPDs using a novel approach.

\section{Methodology}

\begin{figure}[ht]
\centering
\includegraphics[width=1.0\linewidth]{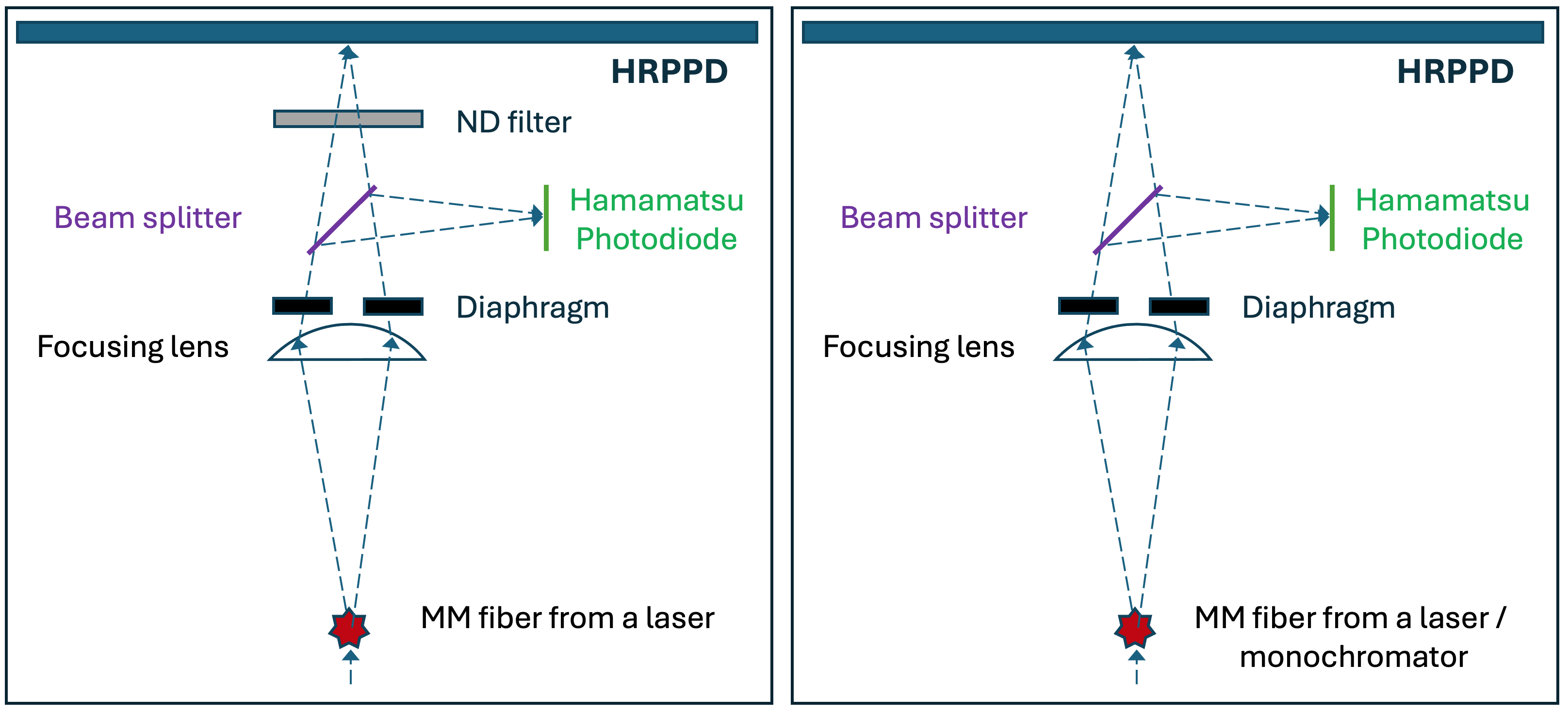}
\caption{A schematic view of our PDE measurement setup (left) and a QE measurement setup (right). See text for details.} \label{fig:1}
\end{figure}

To measure the PDE, we used picosecond diode laser light pulses in a configuration shown in Figure~\ref{fig:1} (left).

The laser beam was split into two paths: direct light focused onto the HRPPD photocathode, at a location matching the center of one of the anode pads (pixels), and light reflected at 90 degrees impinging on a reference photodiode. While a high photon flux is desirable for accurate current monitoring by this photodiode, a much reduced photon flux is required for the HRPPD to ensure a low photon counting mode. To resolve this discrepancy, a neutral density (ND) filter was placed between the beam splitter and the HRPPD. 

Assume that the number of photons incident on the HRPPD in a single laser pulse follows a Poisson distribution with a mean $\lambda$, and that each photon is detected independently with a probability $p$, corresponding to the PDE.
Then, the probability that the HRPPD does not detect any signal from a given pulse is: 
\begin{equation}
P(\text{no signal}) = \sum_{n=0}^\infty P(n) (1-p)^n
\end{equation}
where $P(n) = \frac{\lambda^n e^{-\lambda}}{n!}$.
\begin{equation}
 P(\text{no signal}) = \sum_{n=0}^\infty \frac{(\lambda(1-p))^n e^{-\lambda}}{n!} =  e^{\lambda(1-p)} \times e^{-\lambda} = e^{-\lambda p}   
\end{equation}
\begin{equation}
p= \frac{ln(P(\text{no signal}))}{-\lambda}
\label{MasterFormula}
\end{equation}

Note that the mean photon number $\lambda$ does not need to be small for this expression to hold. In other words, operation in a single-photon mode is not necessary for this method.

The mean number of photons in a single laser pulse ($\lambda$) incident on HRPPD can be derived from the following expression: 
\begin{equation}
\lambda = \frac{I \cdot R}{e \cdot QE \cdot f \cdot A}
\end{equation}
where $I$ is the current as measured by the reference Hamamatsu photodiode (see Figure \ref{fig:1}), QE is its quantum efficiency at the laser wavelength, $R$ is a ratio of the transmitted and the reflected light after the beam splitter at this wavelength, $e$ is the elementary charge, $f$ is a laser repetition rate, and $A$ is an attenuation factor of the ND filter.

Since neither the beam splitter ratio $R$ nor QE of the reference photodiode is known precisely, the $R/QE$ ratio was measured by a Newport NIST traceable photodiode installed in place of an HRPPD, as explained in detail in Section \ref{sec:photodiode-calibration}. The attenuation factor $A$ of the ND filter was also measured separately.
Once the $R/QE$ ratio and the $A$ factor are calibrated, Formula \ref{MasterFormula} gives an estimate of PDE. 

We also evaluated the QE at the same HRPPD active area, by measuring the photocurrent at the same photocathode voltage (see Figure \ref{fig:1} right), and extracted the Collection Efficiency (CE) as a PDE/QE ratio.

It is worth noting that the photon detection efficiency (PDE) is typically determined using a different method~\cite{Lehmann}. In that approach, the pulse amplitude at the anode is measured and then divided by the gain, obtained from a spectral fit, to estimate the number of photoelectrons generated after the photocathode. In contrast, we count signal pulses in a binary manner, thereby eliminating the systematics associated with the spectral fit.

\section{Experimental setup}
\begin{figure}[ht]
\centering
\includegraphics[width=\linewidth]{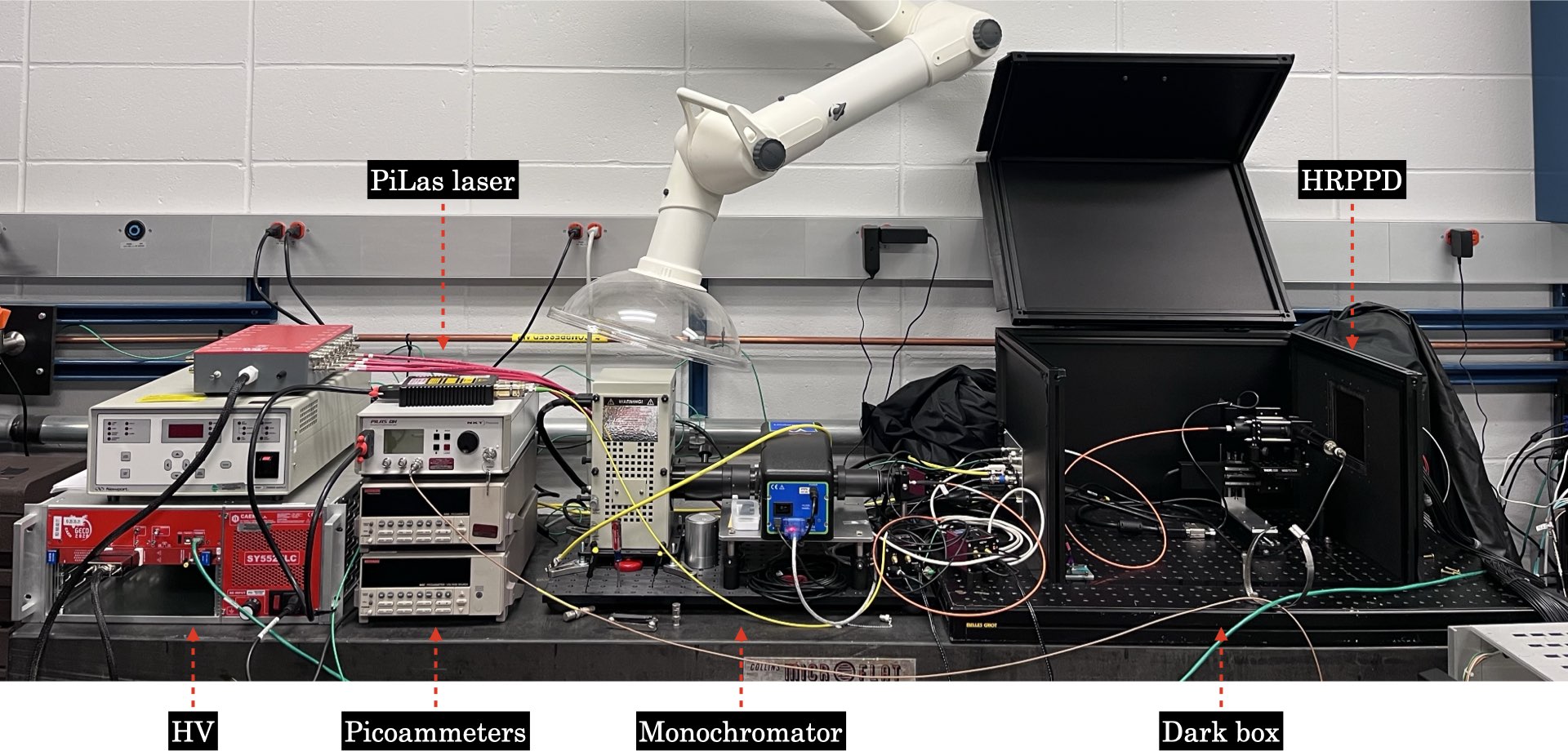}
\caption{Experimental setup.} \label{fig:2}
\end{figure}

A HRPPD test stand at Brookhaven National Laboratory (see Figure \ref{fig:2}) was used to perform the measurements. The equipment, together with the typical settings, is described below. 

\subsection{HRPPD and readout backplane} 

\begin{figure}[ht]
\centering
\includegraphics[width=\linewidth]{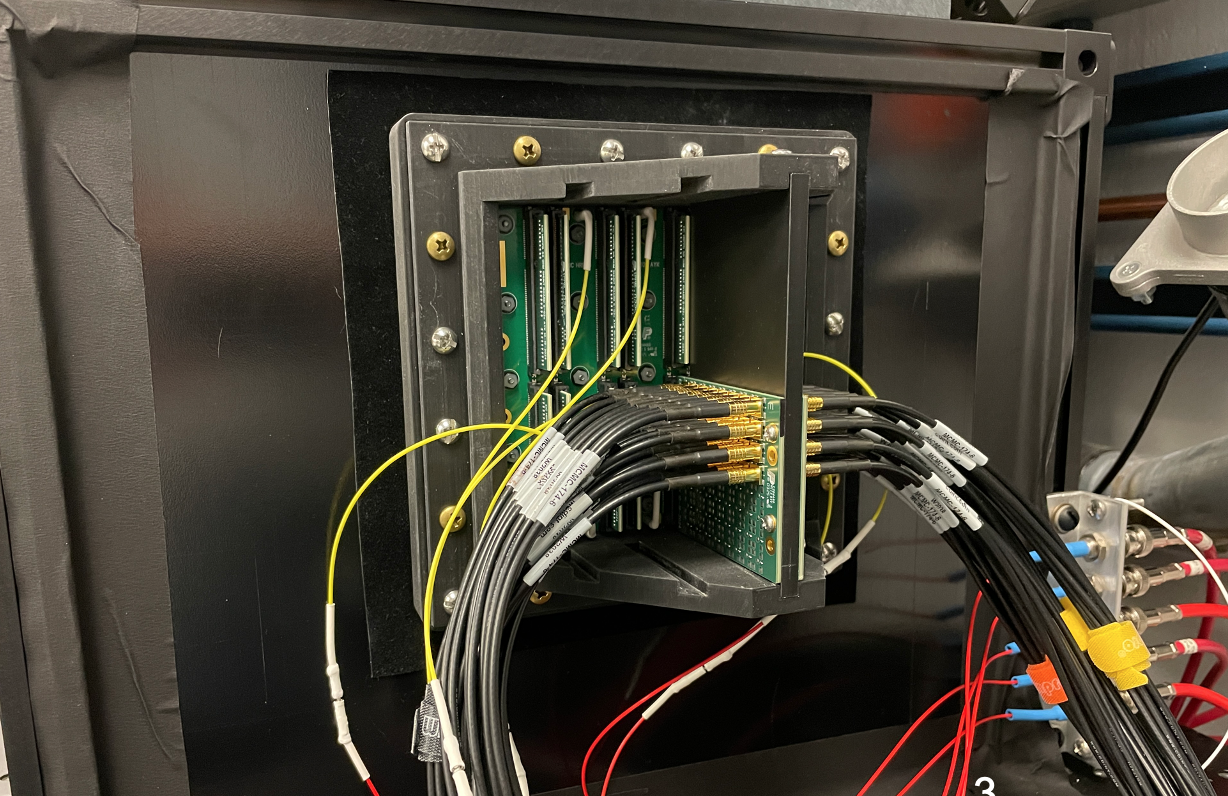}
\caption{HRPPD \#27 mounted on a dark box side wall. One can also see the interface board with a pair of plugin MCX adapters.} 
\label{fig:3}
\end{figure}

    We used HRPPD \#27 (serial number assigned by Incom Inc.) for these measurements. It was mounted on a side wall of the dark box, as shown in Figure \ref{fig:3}. The HRPPD was equipped with a passive readout backplane and a pair of 32-channel MCX adapters, which allowed us to fully instrument a single 8x8 anode pad spot. The HRPPD consists of 16 such 8×8 pad spots, corresponding to a total of 1024 (32×32) channels. For this study, the laser was directed onto a pixel near the center of the active area, from which all results reported in this paper were obtained.
Pads of the other fifteen 8x8 pad spots were terminated to ground via 50-ohm resistors using plugin cards.

\subsection{Dark box equipment}
\label{sec:dark-box}

\begin{figure}[ht]
\centering
\includegraphics[width=\linewidth]{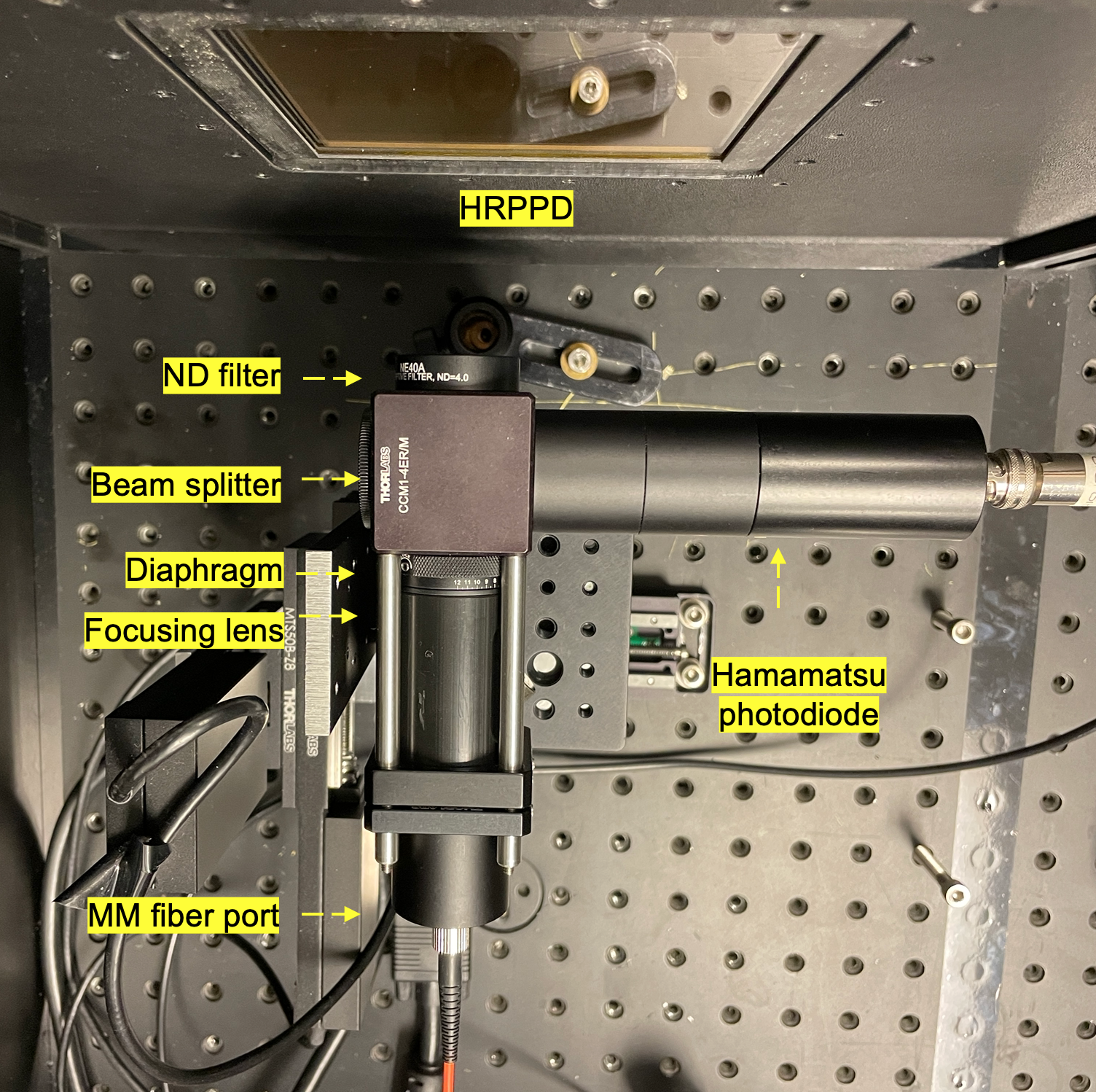}
\caption{A small breadboard with focusing optics mounted on a 3D motion control system inside the dark box. The incoming laser beam, after passing through a beam splitter, was partly focused on the HRPPD photocathode, and partly detected by a reference photodiode. See text for more details.} 
\label{fig:4}
\end{figure}

An optical setup shown schematically in Figure \ref{fig:1} was mounted on an XYZ motion control system (three Thorlabs MTS50/M translation stages) as shown in Figure \ref{fig:4}. 

A UV-resilient UM22-600 multi-mode fiber with a core diameter of 600~$\mu$m by Thorlabs was used between the dark box patch panel and the optical head for both PDE and QE, as well as for the calibration measurements. The light was focused by a Thorlabs LA1422 N-BK7 plano-convex lens with a focal distance of 50~mm, with its intensity and angular spread limited by a Thorlabs SM1D12C diaphragm with a chosen $\sim$6~mm diameter setting, and split into two paths by a 1~mm thick sapphire plate by MSE Supplies used as an asymmetric (about 83:17) beam splitter with a smooth wavelength dependency.

Light reflected at 90 degrees was detected by a Hamamatsu S1226-8BQ reference photodiode installed in a custom support inside a 1" diameter tube, at a distance roughly the same as the distance between the beam splitter and the HRPPD window. 

Transmitted light was detected by either an HRPPD during the measurements (as shown in the picture) or by a NIST traceable Newport 818-UV-DB photodiode for calibration purposes. According to the factory test report, the photodiode responsivity was 1.6864E-1 A/W at 390~nm and 1.7957E-1 A/W at 400~nm in a configuration without an external OD3 filter and 4.0227E-4 A/W at 390~nm and 4.2081E-4 A/W at 400~nm with the filter installed. The uncertainty of this calibration is supposed to be 1.65\% at the wavelength of interest (398.6~nm, see section \ref{sec:laser} below). After a simple linear interpolation,  conversion to quantum efficiency units, and error propagation, we obtain a QE of the Newport photodiode at 398.6~nm to be equal to 55.4 $\pm$ 0.9~\%, and the attenuation factor of the factory OD3 filter at this wavelength equal to 426 $\pm$ 10.

All elements shown in Figure \ref{fig:4} remain unaltered, except the ND filter between PDE and QE measurements. In particular, we used the same fiber, diaphragm diameter setting, and XYZ-coordinates of the optical head. For calibration, either an ND filter or a Newport photodiode was mounted downstream of the beam splitter as explained in the subsequent sections.

\subsection{High Voltage}

For the PDE measurements, the HRPPD was biased using a stackable CAEN A1515BV high voltage power supply in a SY5527LC mainframe. We used the following settings: 600~V bias voltage on each MCP, 200~V for photocathode and transfer voltages. Using this setting, a gain around $5 \times 10^6$ for a particular illuminated pixel is obtained, see also a blue dashed line in Figure \ref{fig:10}. 

For the QE measurements, a 200~V bias was provided by the same Keithley 6487 picoammeter used to measure the photocurrent. Other HRPPD electrodes were floating in this case.
 

\subsection{Laser}
\label{sec:laser}

An NKT Photonics picosecond laser with a PiL1-040-40FC head was used for both the PDE (in a photoelectron pulse counting mode) and the QE (in a photocathode current measurement mode) evaluation.
According to its test report, the laser produced light with a wavelength of 398.6~nm. This value is used in all of our estimates. Laser light was coupled into a single-mode fiber with a nominal core diameter of 3~$\mu$m, and then into a 600~$\mu$m diameter multi-mode fiber inside of the dark box.

\subsection{Monochromator}

A Newport MKS CS130B-1-FH monochromator with a 150~W 66477-150XV-R1 Xenon arc lamp was used for complementary QE measurements in the same setup, to observe a possible difference between a pulsed and a continuous light sources. The monochromator grating grid setting was tuned in such a way that its average wavelength, as measured by a spectrometer (see Section \ref{sec:spectrometer}), was the same as for the laser. 

\subsection{Spectrometer}
\label{sec:spectrometer}

\begin{figure}[ht]
\centering
\includegraphics[width=\linewidth]{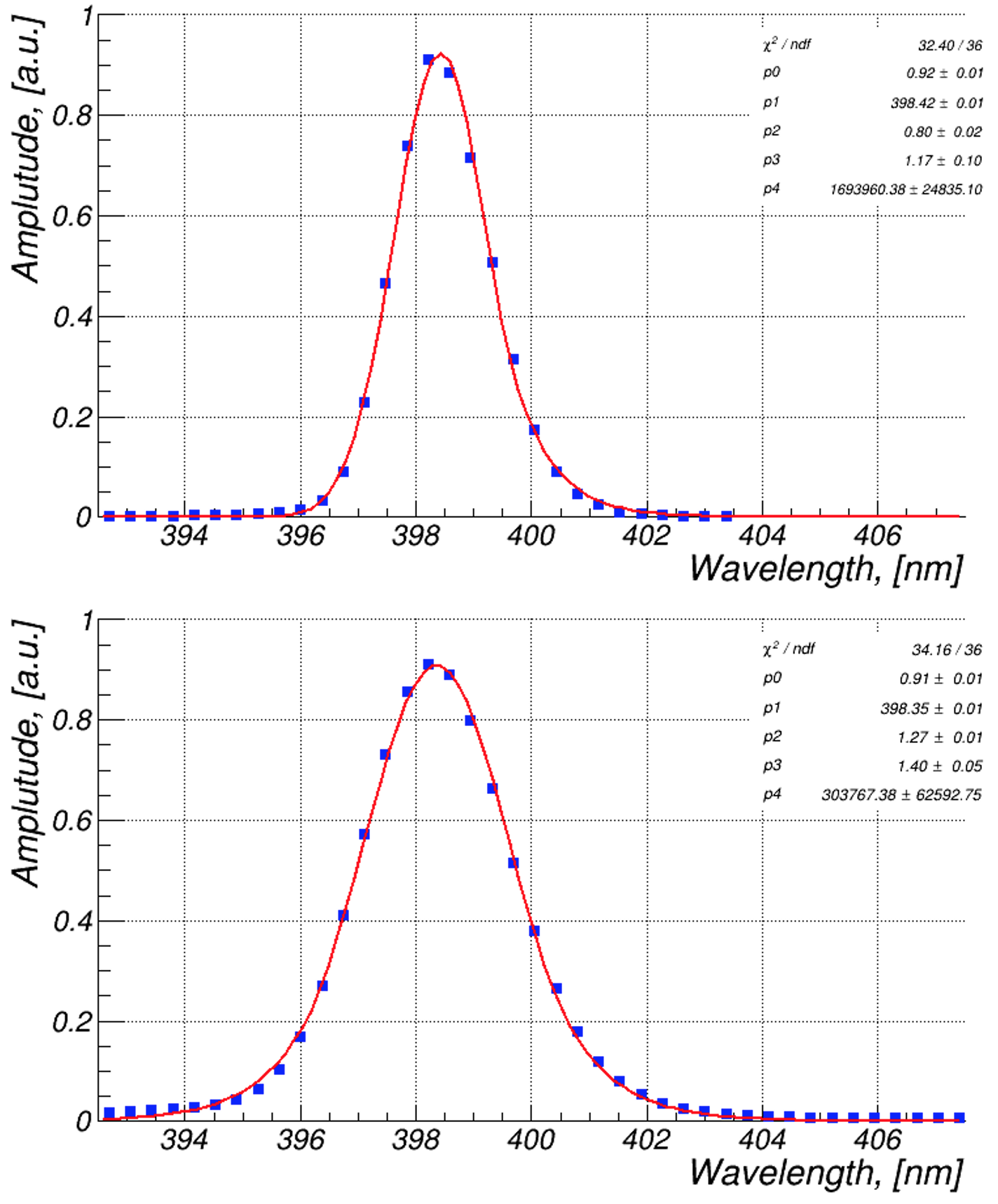}
\caption{Top: PiLas laser wavelength spectrum with a peak value of 398.4~nm and FWHM 1.9~nm as measured by the Ocean Optics spectrometer. Bottom: Newport monochromator wavelength spectrum with a FWHM 3.0~nm and a peak value matching the laser one within 0.1~nm as measured by the spectrometer when using a monochromator wavelength setting of 395.7~nm .}
\label{fig:5}
\end{figure}

A compact SR-6UVV240-25 fiber spectrometer by Ocean Optics was used to measure the actual laser and monochromator wavelengths, as shown in Figure \ref{fig:5}. 

We did not observe any noticeable shift in the laser wavelength in the range of repetition rates between 1~kHz and 40~MHz at either a 0\% or a 60\% tune, as well as in a range of tunes between 0\% and 60\% at a fixed repetition rate of 500~kHz, although the FWHM increased by up to $\sim$10\% at higher tune values.

As follows from Figure \ref{fig:5} (top), a measured PiLas laser wavelength of 398.4~nm is consistent with 398.6~nm provided in the factory report.
However, the monochromator scale was found to be off by almost 3~nm. In the complementary QE measurements described in Section \ref{sec:QE}, we therefore used a slightly different setting of 395.7~nm for the monochromator, in order to match the wavelength spectra peak values between the PiLas laser and the monochromator, as seen in Figure \ref{fig:5}. 
 
\subsection{Picoammeters}

A Keithley 6485 unit was used to measure the photocurrent of the Hamamatsu reference photodiode. A Keithley 6487 was used for the NIST traceable Newport photodiode as well as the HRPPD. 
Data reading is configured in ``SLOW" mode, which corresponds to an integration time of one power line cycle ($\sim$ 16.7 ms).

\subsection{Beam spot imaging}

\begin{figure}[ht]
\centering
\includegraphics[width=\linewidth]{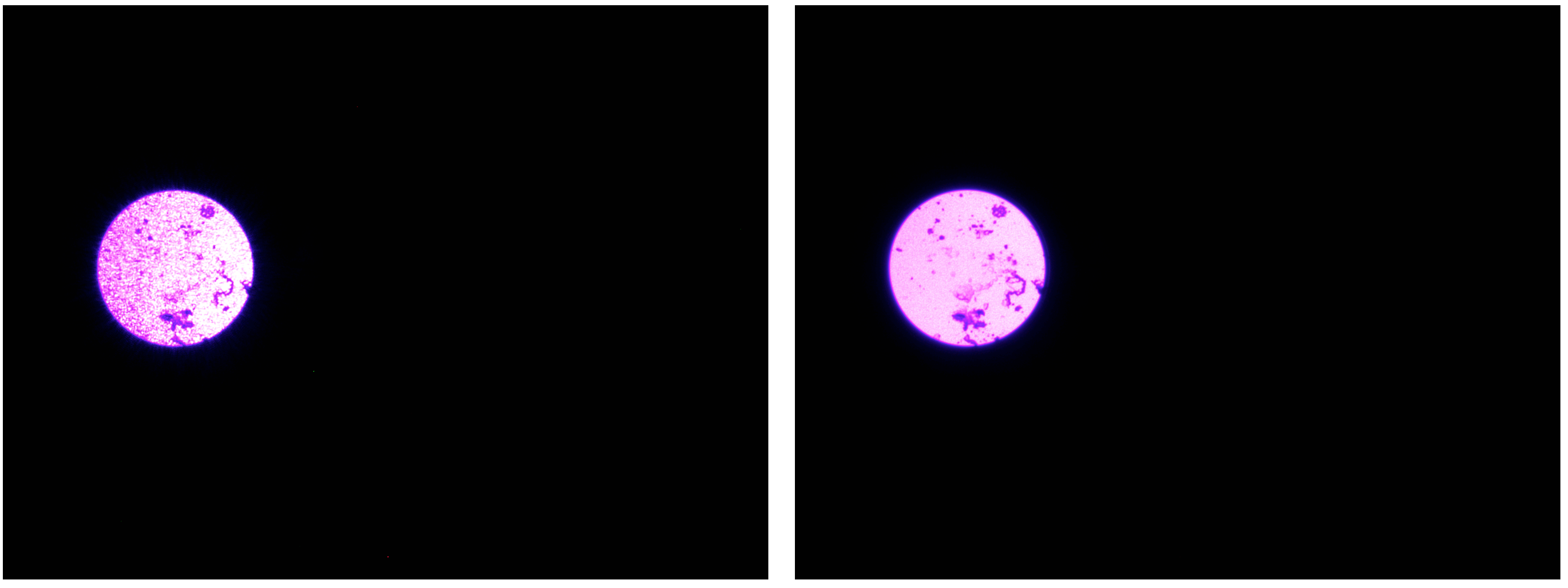}
\caption{Left: PiLas laser beam spot image at the HRPPD window to air boundary location. Right: Newport monochromator beam image in the same optical configuration (fiber type, lens location, diaphragm opening). Sensor dimension: 5.12~mm x 3.84~mm. } 
\label{fig:6}
\end{figure}

We used an AmScope MD310C-BS bare sensor CMOS camera with a 5.12~mm x 3.84~mm sensor  
(2.5~$\mu$m pixels), installed at a location of the HRPPD window to air boundary, to verify the focused beam spot size. Images for both the laser 
and the monochromator light sources are shown in
Figure \ref{fig:6}. As seen from this figure, beam diameter in both cases was about 1.0~mm (mostly defined by the fiber diameter of 600~$\mu$m), though the monochromator image appears more uniform. A beam spot of this size was well within the 8~mm~x~8~mm active area of the Hamamatsu reference photodiode.

\subsection{DAQ, readout electronics, steering software}

Software used to steer all of the setup components, as well as the RCDAQ data acquisition system \cite{RCDAQ} was installed on a Debian 12 desktop PC. Python scripts to operate the monochromator, picoammeters, and translation stages, were written using the {\it PySerial} library. Trigger logic was assembled in a NIM crate, while a pair of CAEN V1742 digitizers in a VME crate was operated via respective CAENET optical links by RCDAQ via a PCIe A3818 interface card.  

\section{Reference photodiode and ND filter calibration}
\label{sec:photodiode-calibration}

\begin{figure}[ht]
\centering
\includegraphics[width=\linewidth]{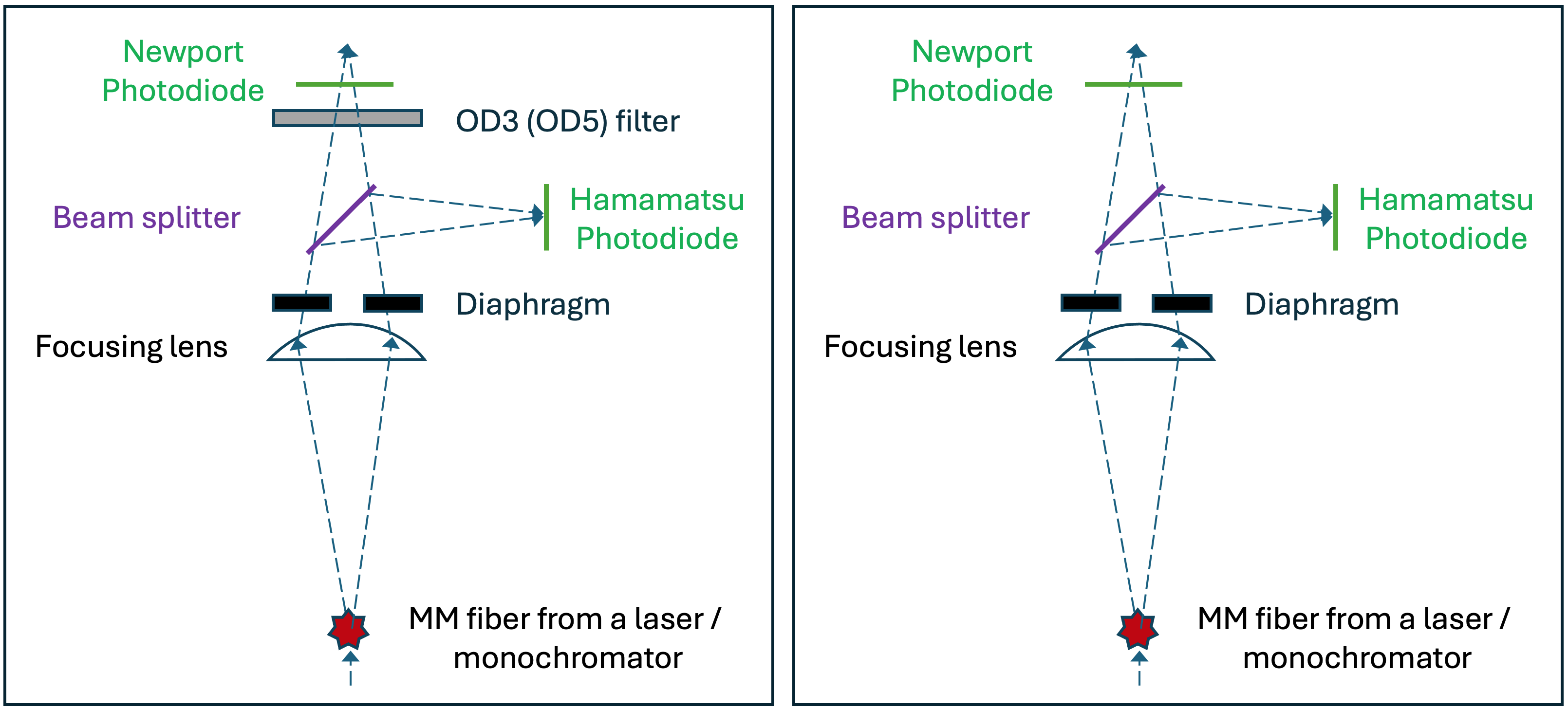}
\caption{Configurations used to cross-calibrate Newport and Hamamatsu photodiodes, as well as our OD5 neutral density filter.} 
\label{fig:7}
\end{figure}

The reference Hamamatsu photodiode does not come with a reliable factory calibration. Besides this, as it was mentioned earlier, neither the beam splitter ratio nor the ND filter optical density at a laser wavelength is known with a sufficient accuracy. In order to relate a reference Hamamatsu photodiode current to a NIST traceable Newport photodiode one, in a configuration with a beam splitter and (optionally) an ND filter installed downstream of it, we performed a series of measurements in configurations shown schematically in Figure \ref{fig:7}. We considered three cases: (1) no filter installed between the beam splitter and the Newport photodiode, (2) an OD3 reflective filter provided with the Newport photodiode (see Section \ref{sec:dark-box}), (3) an OD5 absorptive Thorlabs filter used to perform the PDE measurements.  

\begin{figure}[ht]
\centering
\includegraphics[width=\linewidth]{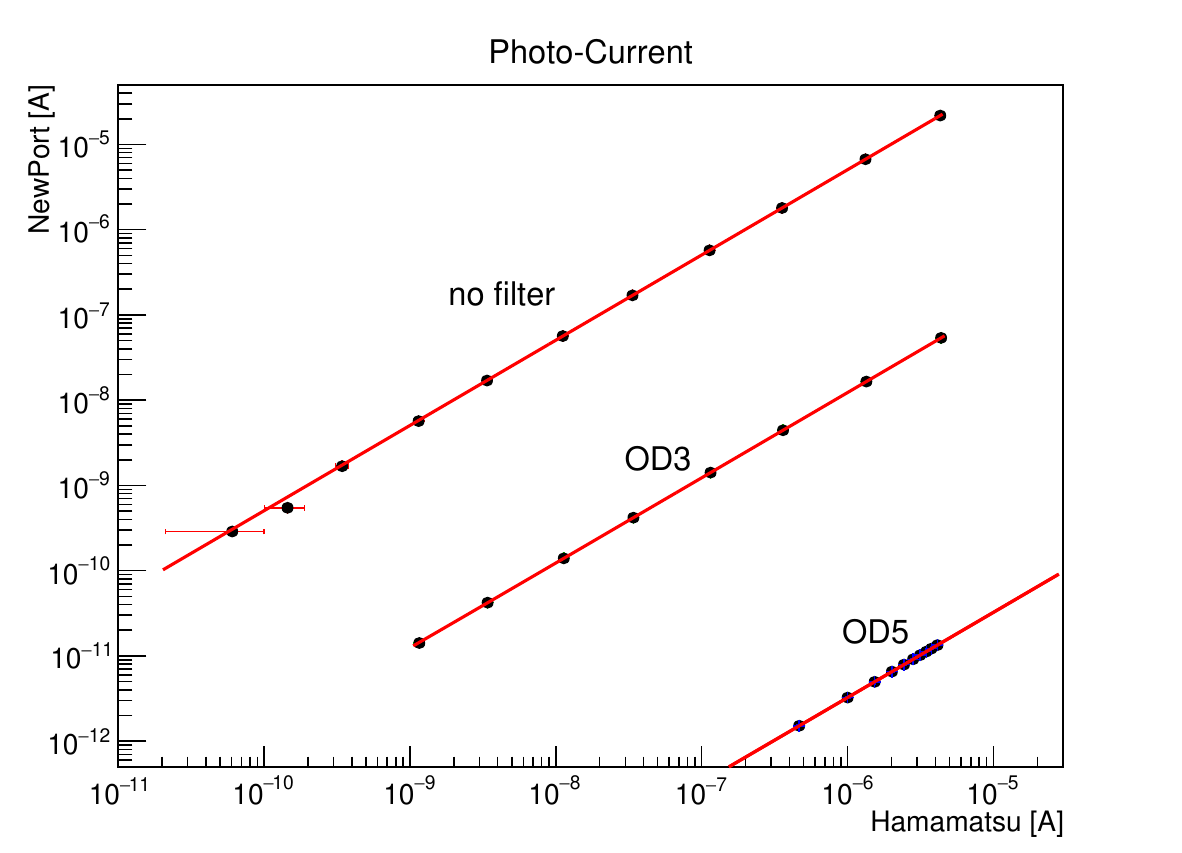}
\caption{Currents measured by the Newport and Hamamatsu photodiodes at various PiLas laser light intensities, for the configurations shown in Figure \ref{fig:7}. The linear fit for each configuration is shown.} 
\label{fig:8}
\end{figure}

Results are presented in Figure \ref{fig:8}. The PiLas Laser was set at a 0\% tune (maximal pulse intensity). Average integrated light intensity was varied by changing a laser repetition rate decreasing from 40~MHz in roughly half a decade steps (11 points on a ``no filter" curve in Figure \ref{fig:8}). Ten current measurements per point were taken, and their average and standard deviation are used in three separate $f(x) = ax+b$ straight line fits. In all three cases, we observed a linear response within the measurement errors ($\chi^2$ of 3.12, 0.54, and 0.39 per 9, 6, and 8 degrees of freedom, respectively), and $b$-terms consistent with 0. (Note that the small values of $\chi^2$ are attributed to the overestimated uncertainties.)

We make the following observations: 
\begin{enumerate}[label=(\roman*)]
\item Hamamatsu photodiode has a linear response over four decades in light intensity. 
\item One can use a slope of the ``no filter" straight line in Figure \ref{fig:8} to calibrate the Hamamatsu reference photodiode against the Newport one for QE measurements.
\item In a similar way, one can use a ratio of the ``OD5 filter" and ``no filter" line slopes in Figure \ref{fig:8} to cross-calibrate the Hamamatsu photodiode in a low light intensity configuration required to measure the HRPPD PDE in a photoelectron counting mode. 
\end{enumerate}

Numerically, the slope of the ``no filter" straight line was found equal to 5.0416 $\pm$ 0.0011, and the ratio of the ``no filter" and ``OD5 filter" slopes (effectively the optical density of the OD5 filter at 398.6~nm) was equal to 
(1556 $\pm$ 5)E3.

It is worth mentioning that the ``Newport OD3 filter" data were taken with a neutral density filter, for which
a separate factory calibration was provided (see Section \ref{sec:dark-box}). Our measured slope ratio of the ``Newport OD3 filter" and ``no filter" curves leads to an attenuation factor of 411.0 $\pm$ 0.2, consistent with these factory calibrations at 398.6~nm within the quoted errors.

\section{HRPPD PDE measurement}

\begin{figure}[ht]
\centering
\includegraphics[width=\linewidth]{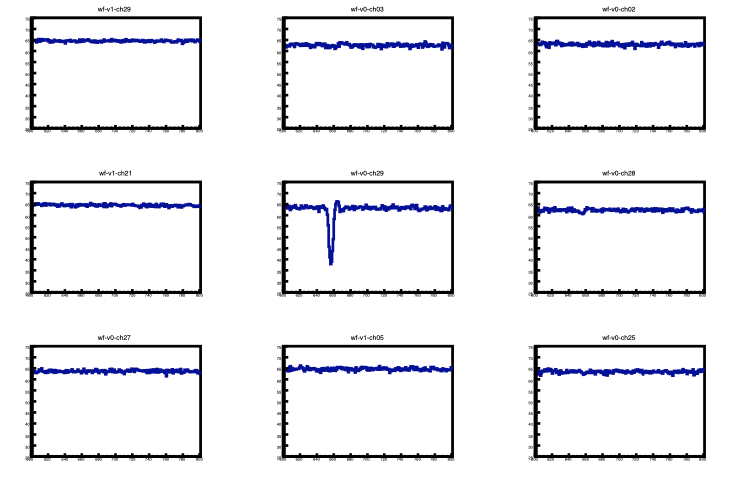}
\caption{A single event with waveforms of the 3x3 HRPPD pad signals around the illuminated one. The horizontal range is 40~ns and the vertical range is 50~mV in each of the waveforms of the 3x3 pads surrounding the illuminated one. A single-electron pulse is observed in the illuminated channel.} 
\label{fig:9}
\end{figure}

The measurement was performed in a configuration shown schematically in Figure \ref{fig:1}, with the OD5 neutral density filter installed between the beam splitter and the HRPPD window. The PiLas laser was run with a repetition rate of 1~kHz at a set of tunes listed in Table \ref{tab:result-pde}. The RCDAQ DAQ was used to take the waveform data of all the 8x8 pads around the illuminated spot. The V1742 digitizers were running at a 5~GS/s sampling rate. The trigger for the DAQ was provided by the PiLas laser itself. A panel of a typical ``event" is shown in Figure \ref{fig:9}. 

\begin{figure}[ht]
\centering
\includegraphics[width=\linewidth]{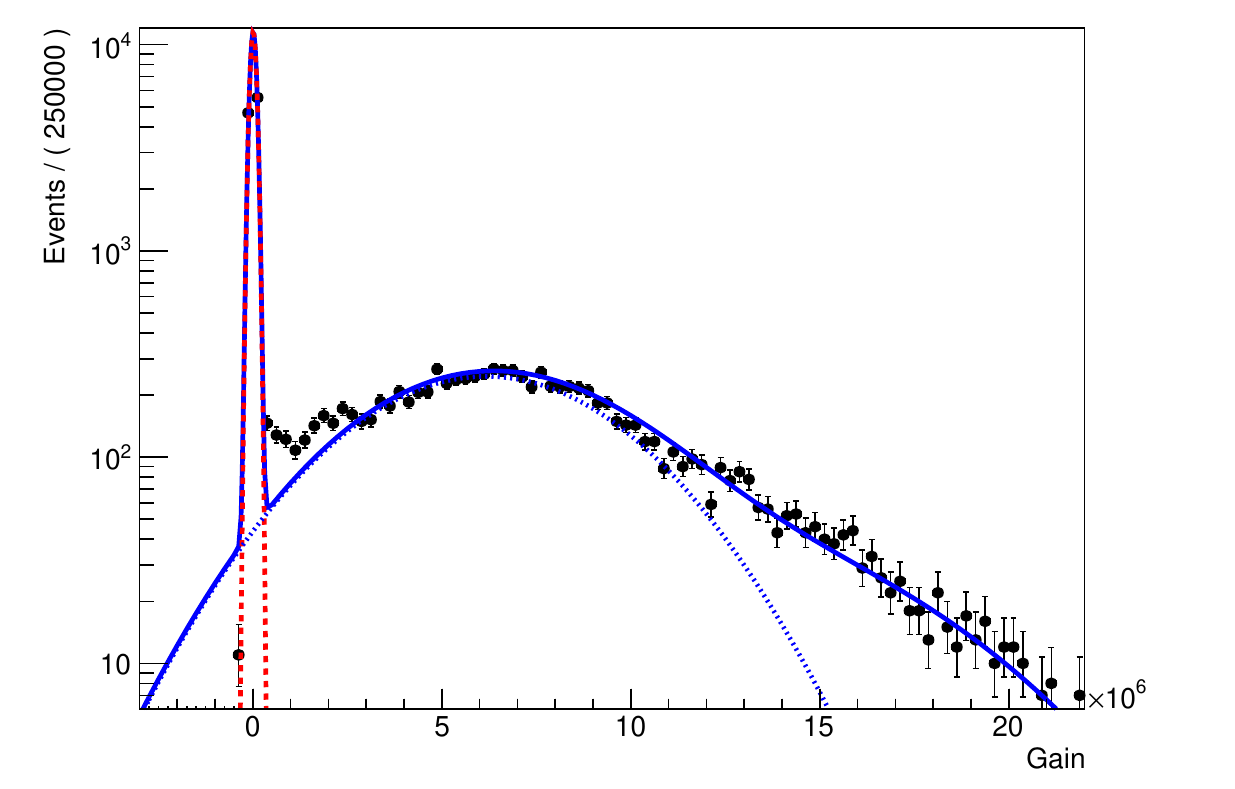}
\caption{Charge collected by the illuminated HRPPD pad. Red dashed line indicates a fit to the pedestal, blue dashed line the one-photon peak. Blue solid line is the sum of the fits to one-photon peak and two-photon peak.} 
\label{fig:10}
\end{figure}

Despite the fact that a majority of recorded events looked like in Figure \ref{fig:9} (only the illuminated pad had a measurable signal), various summation schemes for a 3x3 pad area around the illuminated spot have been tested. Ultimately, we decided to use waveforms of the illuminated pad and four neighboring pads for the collected charge measurements, which contain most part ($>$95\%) of the signal charge and provide a relatively narrow pedestal peak. No peak search algorithm was used. The HRPPD signal was taken as a sum of 16 waveform samples (a 3.2~ns wide window) at a fixed timing offset with respect to a digitized trigger pulse. The baseline was estimated on an event-per-event basis in a same width (3.2~ns) window just in front of the location of the expected HRPPD peak, in the same waveform. Collected charge was calculated as a difference between the ``peak" value (a sum over 16 samples as explained earlier, whether there was a measurable peak observed or not) and a baseline. Charge distribution, normalized to a detected electron count (which in an ideal single photon mode would effectively have a meaning of the HRPPD gain) is shown in Figure \ref{fig:10}, together with a Gaussian fit to a pedestal peak (events where HRPPD did not produce a measurable signal) and a two-Gaussian fit to the actual signal distribution, assuming that three-photoelectron event probability was small and therefore irrelevant.   

One can see that the HRPPD was operated at a gain of about $5 \times 10^6$. The pedestal peak had $\sigma \sim 0.09 \times 10^6$.
A fraction of events with a measurable HRPPD signal required in formula \ref{MasterFormula} was determined as a ratio of events with a collected charge above $3\sigma$ with respect to the pedestal mean, and the total number of triggers.

Since the Hamamatsu reference photodiode current was recorded at the same time, both in LASER ON and LASER OFF configurations to subtract pedestal, an average number of photons impinging onto the HRPPD photocathode per laser pulse was known precisely. Substituting the numbers into formula \ref{MasterFormula}, and propagating the statistical errors, one obtains the PDE values.

We repeated the same measurements at a set of different laser tunes corresponding to a substantially different average fraction of non-empty events, and obtained consistent PDE values, all listed in Table~\ref{tab:result-pde}. A statistical average of (17.13 $\pm$ 0.11)\% ($\chi^2/$NDF = 0.87/4, p-value = 0.9283) is taken as a final result.

\begin{table}[ht]
\centering
\begin{tabular}{c|c|c|c}
\toprule
tune & $\lambda$ & 
signal / total &
PDE \\
\midrule
0\%    & 3.962$\pm$0.007   
& (49.09$\pm$0.35)\% & (17.04$\pm$0.22)\%  \\
10\%   & 3.368$\pm$0.007   
& (43.99$\pm$0.35)\% & (17.21$\pm$0.23)\%  \\
15\%   & 2.968$\pm$0.008 
& (40.13$\pm$0.35)\% & (17.28$\pm$0.24)\%  \\
25\%   & 2.300$\pm$0.005 
& (32.42$\pm$0.33)\% & (17.03$\pm$0.23)\%  \\
40\%   & 1.678$\pm$0.006 
& (24.95$\pm$0.31)\% & (17.11$\pm$0.26)\%  \\
\bottomrule
\end{tabular}
\caption{Results of five sets of data using different laser tunes, listed in the first column. The second column ($\lambda$) is the mean number of photons impinging on the HRPPD photocathode per laser pulse. Only statistical errors are quoted.}
\label{tab:result-pde}
\end{table}

\subsection{Statistical and systematic uncertainties}

Statistical uncertainties quoted in Table \ref{tab:result-pde} follow from 
the spread of the reference photodiode current measurements and binomial error when calculating a fraction of events that have HRPPD pulses above threshold with its error propagation in formula \ref{MasterFormula}.

The following sources of systematic uncertainty are taken into account: a 1.65\% relative uncertainty of the Newport photodiode calibration, and a 0.32\% relative uncertainty of the attenuation factor of the OD5 filter from the fit errors of the photodiode calibration curves (see Section \ref{sec:photodiode-calibration}).

\section{HRPPD QE measurement}
\label{sec:QE}

We performed the QE measurement using both the PiLas laser  (at a variety of repetition rates and tunes) and the monochromator with xenon lamp, in order to see a possible difference between a pulsed and a continuous light source. A configuration shown in Figure \ref{fig:1} (right) was used in both cases, without changing anything inside of the dark box. The photocathode bias voltage was 200~V, provided by the Keithley 6487 picoammeter.
The results are listed in Table \ref{tab:result-qe}, where uncertainties for QE come from three parts: the fluctuation of the photocurrent of the HRPPD photocathode, the fluctuation of the photocurrent of the reference photodiode, and the fit error of the ``no filter" curve in Figure~\ref{fig:8}. All measurements were carried out after the HV on HRPPD had been turned on for 3 hours. The light source was turned off for about ten minutes between the successive measurements.

\begin{table}[ht]
\centering
\begin{tabular}{l|c|c}
\toprule
light source  &
{\small HRPPD photocurrent} & QE\\
\midrule
{\small Laser @10~MHz tune 0\%}     & 3043$\pm$35~nA& (22.88$\pm$0.26)\% \\ 
{\small Laser @40~MHz tune 98\%}     & 1069$\pm$13~nA& (22.95$\pm$0.46)\% \\ 
{\small Laser @1~MHz tune 0\%}     & 267.5$\pm$0.7~nA& (23.38$\pm$0.06)\% \\ 
{\small Laser @0.5~MHz tune 0\%}     & 132.0$\pm$0.2~nA& (23.67$\pm$0.04)\% \\
{\small Laser @0.1~MHz tune 0\%}     & 26.46$\pm$0.01~nA& (23.76$\pm$0.01)\% \\ 
{\small Laser @50~kHz tune 0\%}     & 13.23$\pm$0.02~nA& (24.37$\pm$0.05)\% \\ 
{\small Laser @30~kHz tune 0\%}     & 7.99$\pm$0.01~nA& (24.37$\pm$0.03)\% \\ 
Monochromator                       & 231.6$\pm$2.3~nA& (24.05$\pm$0.39)\% \\ 
\bottomrule
\end{tabular}
\caption{HRPPD QE measurement results. Only statistical errors quoted.}
\label{tab:result-qe}
\end{table}

One can see that the measured values exhibit a certain trend in the laser data, specifically that the QE estimates increase with decreasing HRPPD photocathode currents. We stop at 30 kHz since the photocurrent has decreased to a similar magnitude of the leakage current ($\sim$6~nA) and the result remained identical to what we obtained at 50 kHz. 

We are therefore taking a QE value of (24.37 $\pm$ 0.03 [stat])\%, obtained with a PiLas laser at the lowest intensity presented in Table \ref{tab:result-qe}, as final result with a substantial systematic uncertainty 0.32\% taken from the difference between the laser measurement and the monochromator measurement. As a cross-check, the QE and PDE of a nearby pixel on this HRPPD, as well as of a pixel on another HRPPD, were also measured, yielding consistent results. It is worth noting that uniformities of QE, and consequently of PDE, can exhibit a relative variance exceeding 20\% across the entire active area of the HRPPD~\cite{Lyashenko}.

\section{HRPPD Collection Efficiency}

\begin{figure}[ht]
\centering
\includegraphics[width=0.8\linewidth]{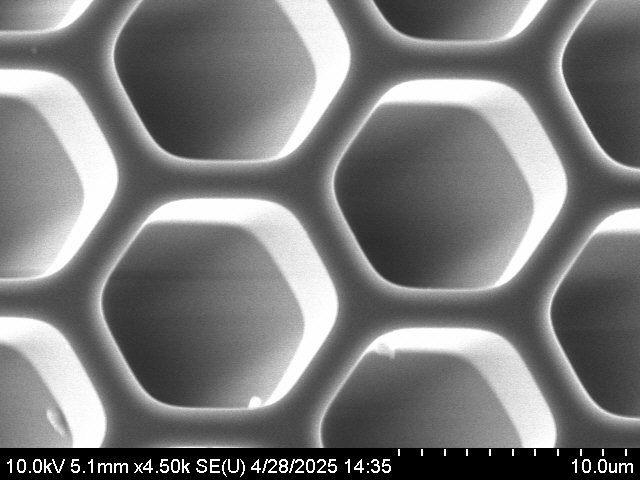}
\caption{An MCP sample image from the same capillary glass array block as used for manufacturing the HRPPD \#27, taken by a scanning electron microscope, shows the configuration of the $\sim10\mu m$ diameter pores and the interstitial space between them.} 
\label{SEM}
\end{figure}

The Collection Efficiency follows from the PDE and QE measurements at the same wavelength and at the same physical location in the HRPPD active area as CE = PDE/QE. Substituting the numbers obtained in the last two sections, one can see that the HRPPD CE is (70.3 $\pm$ 1.6)\%, as measured at a particular active area spot. This value is reasonable given that most of the primary photoelectrons striking the interstitial space between the pores of the top surface of the first MCP (Figure~\ref{SEM}) either do not bounce back or do not make it into the pores with an energy sufficient to initiate an avalanche. For the MCPs used in EIC HRPPD \#27, its open area ratio of the entry surface is 69.4\%~\cite{OAR}, in a reasonable agreement with our CE estimate. Images from scanning electron microscopy reveal that the open area ratio across the MCP entry surface varies by at most 4\% for a beam spot with 1~mm diameter.

\section{Conclusions}
The photon detection efficiency of one HRPPD from the first batch dedicated for EIC has been measured to be (17.1~$\pm$~0.1~[stat]~$\pm$~0.3~[sys])\% at 398.6~nm, for a pixel near the center, in a photoelectron pulse counting mode using a picosecond diode laser. HRPPD QE at the same spot and at the same wavelength has been evaluated to be (24.4~$\pm$~0.1~[stat]~$\pm$~0.3~[sys])\%, leading to a CE estimate of (70.3 $\pm$ 1.6)\%, which is consistent with the expectations. 



\section*{Acknowledgements}
We are grateful to Dr.~Zhoudunming Tu and the Center for Functional Nanomaterials at BNL for their assistance with the scanning electron microscope imaging and to Bob Azmoun for his contribution to the setup commissioning. This work was supported in part by
the U.S. Department of Energy under Prime Contract No. DE-SC0012704.


\bibliography{Bibliography.bib}




\end{document}